\documentclass[a4paper,12pt]{article}
\usepackage[cmex10]{amsmath}
\usepackage{latexsym,amssymb}
\usepackage[mathscr]{eucal}
\usepackage{amsxtra,amscd,amsthm,upref,layout,color}
\usepackage{calc}
\usepackage[final]{graphicx}
\def\refup#1{{$^{#1}$}}

\def\oo{{\leavevmode\setbox0=\hbox{h}\dimen0=\ht0 \advance\dimen0
by-1ex\rlap{\raise0.47\dimen0\hbox{\char'27}}o}}
\def\begeq{\begin{equation}}
\def\endeq{\end{equation}}
\def\begdis{\begin{displaymath}}
\def\enddis{\end{displaymath}}

\def\cA{{\cal A}}\def\cC{{\cal C}}

\def\Li{{\rm Li}}

\def\ie{{\em i.e.}}\def\eg{{\em e.g.}}

\begin{document}
\title{Thermodynamic equivalence of physical systems}
\author{  
{{Salvino Ciccariello}}\\
  \begin{minipage}[t]{0.9\textwidth}
   \begin{flushleft}
   \setlength{\baselineskip}{12pt}
{\slshape  {\footnotesize{Universit\`{a} di Padova,
   Dipartimento di Fisica {\em G. Galilei}}}}\\
  {\slshape{\footnotesize{Via Marzolo 8, I-35131 Padova, Italy}} }\\
  \footnotesize{salvino.ciccariello@unipd.it}
\end{flushleft}
\end{minipage}
}
\date{\today}
\maketitle                        
\begin{abstract} \noindent 
Two different physical systems are said to be thermodynamically equivalent if 
one of the thermodynamic potentials of the first system is proportional to  the 
corresponding potential of the second system after expressing the state variables 
of the first system in terms of those of the second by a transformation reversible 
throughout the state parameter domain. 
The thermodynamic equivalence has a transitive nature so that physical  
systems divide into classes of thermodynamically equivalent systems that 
have similar phase diagrams. A first class of thermodynamically equivalent 
systems is formed by the ideal classical and quantum Fermi gases, whatever the 
dimensions of the confining spaces, and the one dimensional hard rod gas. 
A second  class is formed 
by the physical systems characterized by interactions that coincide by a scaling 
of the distance and the coupling constant.  {A third class is formed by the ideal 
Boses gases in arbitrary spatial dimensions. The thermodynamic equivalence can 
also be defined in a more general way. By so doing the first and the 
third class combine into a single  class  of thermodynamically equivalent systems.} 
   \\


\end{abstract}
\vfill
\eject
{}{}

\def\kb{{k_{_{B}}}}
\def\Na{{N_A}}\def\Nb{{N_B}}\def\mua{{\mu_A}}\def\mub{{\mu_B}}\def\muf{{\mu_F}}
\def\Vf{{V_F}}\def\Tf{{T_F}}
\def\za{{z_A}}\def\zb{{z_B}}\def\zf{{z_F}}
\def\Ta{{T_A}}\def\Tb{{T_B}}\def\Sa{{S_A}}\def\Sb{{S_B}}
\def\Va{{V_A}}\def\Vb{{V_B}}\def\Vig{V{_{ig}}}\def\Tig{T_{ig}}\def\zig{z_{ig}}\def\pa{{p_A}}
\def\pb{{p_B}}\def\Ua{{U_A}}\def\Fa{{F_A}}\def\Ga{{G_A}}\def\Ha{{H_A}}\def\gpa{{\Omega_A}}
\def\Ub{{U_B}}\def\Fb{{F_B}}\def\Gb{{G_B}}\def\Hb{{H_B}}\def\gpb{{\Omega_B}}
\def\ga{{g_{_A}}}\def\gb{{g_{_B}}}\def\siga{{\sigma_{_A}}}\def\sigb{{\sigma_{_B}}}
\def\ma{{m_{_A}}}\def\mb{{m_{_{B}}}}
\def\muhr{{\mu_{_{hr}}}}\def\nhr{{n_{_{hr}}}}\def\muig{{\mu_{_{ig}}}}\def\nig{{n_{_{ig}}}}
\def\pig{{p_{_{ig}}}}
\def\gphr{{\Omega_{_{hr}}}}\def\gpig{{\Omega_{_{ig}}}}\def\gpigD{{\Omega_{_{ig,\,D}}}}
\def\gpigt{{\Omega_{_{ig,\,3}}}}\def\gpigu{{\Omega_{_{ig,\,1}}}}
\def\lamhr{{\lambda_{_{hr}}}}\def\lamig{{\lambda_{_{ig}}}}\def\lam{{\lambda}}
\def\Nhr{N_{_{hr}}}\def\Nig{N_{_{ig}}}\def\Shr{S_{_{hr}}}\def\Sig{S_{_{ig}}}
\def\Lhr{L_{_{hr}}}\def\Lig{L_{_{ig}}}\def\Thr{T_{_{hr}}}\def\Tig{T_{_{ig}}}
\def\phr{p_{_{hr}}}\def\pig{p_{_{ig}}}\def\sigab{\sigma{_{AB}}}\def\sigba{{\sigma_{_{BA}}}}
\def\Li{{\rm Li}}
\def\gb{{g_{_{B}}}}\def\gf{{g_{_{F}}}}\def\te{{\it  {te}}}
\section{Introduction}  
The thermodynamic behaviour of an open physical system is fully determined by 
the knowledge of one of its  thermodynamic potentials which depends on a triple 
of state  variables (one of which, at least, extensive)  chosen among the three pairs  
of conjugate variables $(V,\,p)$, $(S,\,T)$ and $(N,\,\mu)$. Here,  the extensive 
variables $V,\, S$ and $N$ respectively denote the volume, the entropy and the 
particle number of the sample while the intensive variables $p,\, T$ and $\mu$ 
are the pressure, the temperature and the chemical potential\refup{[1,2]}.  
For instance, choosing $V,\,S$ and $N$ as state variables,  the thermodynamic 
potential that fully determines the thermodynamic behavior of the system is the 
internal energy  $U(V,\,S,\,N)$ equal to 
\begeq\label{1.1}
U(V,\,S,\,N)=-p(V,\,S,\,N)\,V+T(V,\,S,\,N)\,S+\mu(V,\,S,\,N)\,N
\endeq
with 
\begin{eqnarray}
p=p(V,\,S,\,N)&=&-\frac{\partial U(V,\,S,\,N)}{\partial V}\label{1.2},\\
T=T(V,\,S,\,N)&=&\frac{\partial U(V,\,S,\,N)}{\partial S},\label{1.3}\\
\mu=\mu(V,\,S,\,N)&=&\frac{\partial U(V,\,S,\,N)}{\partial N}.\label{1.4}
\end{eqnarray}
If one wishes to determine another thermodynamic potential, say the grand 
potential $\Omega(V,\,T,\,\mu)=-p\,V$, one solves equations (\ref{1.3}) and 
(\ref{1.4}) with respect to $S$ and $N$ so as to get 
\begeq\label{1.5}
S=S(V,\,T,\,\mu)\quad{\rm and}\quad N=N(V,\,T,\,\mu).
\endeq 
After substituting these into equation(\ref{1.2}) one finds the expression 
of the grand potential in terms of its natural variables, \ie 
\begeq\label{1.6}
\Omega(V,\, T,\,\mu)=-p(V,\, T,\,\mu)\,V.
\endeq 
In this way the transformation $(V,\,S,\,N)\to (V,\, T,\,\mu)$, defined by  
\begeq\label{1.7}
V \to V,\quad S\to S(V,\,T,\,\mu)\quad{\rm and}\quad N\to N(V,\,T,\,\mu),
\endeq 
allows one to describe the thermodynamic behaviour of the system using the 
grand potential instead of the internal energy. \\ 
Consider now two different physical systems A and B in thermodynamic equilibrium. 
One wonders on the implications of a reversible transformation  between the state 
variables relevant to a thermodynamic potential of the first system and the state
 variables of the same thermodynamic potential relevant to the second system. 
These transformations lead to the definition of  the thermodynamic equivalence 
(TE) that  in turns unifies the thermodynamic behaviour of different physical systems. 
 { To the author knowledge, Lee\refup{[3]}  first spoke of TE  when he realized that 
the grand potentials of the ideal Fermi and Bose gases are related  to 
the same mathematical functions. Moreover,  after equating the temperatures and 
the volumes of the two gases, Lee\refup{[4,5]} got, in the two dimensional 
case,  the fugacity transformation that converts the grand potential of the Fermi gas into 
that of the Bose gas plus a known function. Some years later, Anghel\refup{[6]} showed 
that the last result also applies to two dimensional systems with particles obeying fractional 
statistics.  This paper exploits the consequences  of  the TE definition 
that, as shown in section 2, can be formulated into a restricted and a generalized 
way. The analysis will be mainly confined to the first definition because this ensures that the 
state diagrams of thermodynamically equivalent (\te) systems are similarly shaped.  
Three examples of different physical systems that are te  are discussed in sections 3 
(ideal classical gases), 4 (systems characterized by a scalable interaction) and 5 
(quantum ideal gases). The last section also shows that two physical systems, which  
are  not \te\  if one adopts the restricted 
definition of TE,  become \te\ with  the  more general definition of TE. } 
Finally, section 6 draws the final conclusions. \\ 
\section{Thermodynamic equivalence}
For definiteness let us refer to the grand potentials  $\gpa(\Va,\,\Ta,\,\mua)$ and  
$\gpb(\Vb,\,\Tb,\,\mub)$ of systems A and B and denote the transformation 
from $(\Vb,\,\Tb,\,\mub)$ of B to $(\Va,\,\Ta,\,\mua)$ of A as    
\begin{eqnarray}
\Vb\to \Va&=&V_{_{A,B}}(\Vb,\,\Tb,\,\mub)\equiv \Vb\,v_{a,b}(\Tb,\,\mub),\label{1.8}\\
\Tb\to \Ta&=&T_{_{A,B}}(\Vb,\,\Tb,\,\mub)\equiv t_{a,b}(\Tb,\,\mub),\label{1.9}\\
\mub\to \mua&=&\mu_{_{A,B}}(\Vb,\,\Tb,\,\mub)\equiv {\bar\mu}_{a,b}(\Tb,\,\mub).\label{1.10}
\end{eqnarray}         
It is noted that the functional forms reported in the rightmost members  
of the above three relations are dictated by the extensive or intensive nature of the 
considered variables. The transformation is required to be endowed of continuous partial 
derivatives at least up to the second order, to be reversible and to be physically 
meaningful. The last condition implies that $v_{a,b}(\Tb,\,\mub)$ and $t_{a,b}(\Tb,\,\mub)$ 
must be strictly positive and the second that the codomains of functions 
(\ref{1.8})-(\ref{1.9}) coincide with the physical ranges of variables $\Va,\,\Ta$ and 
$\mua$.  Denoting the inverse functions of $t_{a,b}(\Tb,\,\mub)$ and 
${\bar\mu}_{a,b}(\Tb,\,\mub)$ as   $t_{b,a}(\Ta,\,\mua)$ and ${\bar\mu}_{b,a}(\Ta,\,\mua)$,  
the inverse transformation of equations  (\ref {1.8})-(\ref{1.10}) reads 
 \begin{eqnarray}
\Va\to \Vb&=&V_{_{B,A}}(\Va,\,\Ta,\,\mua)\equiv \Va\,v_{b,a}(\Ta,\,\mua),\label{1.11}\\
\Ta\to \Tb&=&T_{_{B,A}}(\Va,\,\Ta,\,\mua)\equiv t_{b,a}(\Ta,\,\mua),\label{1.12}\\
\mua\to \mub&=&\mu_{_{B,A}}(\Va,\,\Ta,\,\mua)\equiv {\bar\mu}_{b,a}(\Ta,\,\mua),\label{1.13}
\end{eqnarray}
where  
\begeq\label{1.14}
v_{b,a}(\Ta,\,\mua)\equiv 1/v_{a,b}(\Tb,\,\mub)=1/v_{a,b}(t_{b,a}(\Ta,\mua),\,
{\bar\mu}_{b,a}(\Ta,\,\mua)).
\endeq 
Applying the state variable transformations (\ref{1.8})-({\ref{1.10}}) to the grand potential 
of system A, one obtains the new function $\Omega_{_{A,B}}$ defined as 
\begin{eqnarray}\label{1.14a}
&&\quad \quad \Omega_{_{A,B}}(\Vb,\,\Tb,\,\mub)\equiv \\
 && \gpa\bigl(V_{_{A,B}}(\Vb,\,\Tb,\,\mub),
T_{_{A,B}}(\Vb,\,\Tb,\,\mub),\mu_{_{A,B}}(\Vb,\,\Tb,\,\mub)\bigr).\nonumber
\end{eqnarray}     
Systems A and B are said to be thermodynamically equivalent (\te) if function 
$\Omega_{_{A,B}}$ is proportional to the grand potential of system B, \ie 
\begeq\label{1.15}
\Omega_{_{A,B}}(\Vb,\,\Tb,\,\mub)=C_{_{AB}}\gpb\bigl(\Vb,\,\Tb,\,\mub),
\endeq 
$C_{_{AB}}$ being a real positive constant. \\ 
From this definition immediately follows that the thermodynamic equivalence (TE) may 
occur  if: {\em a)} {\em the codomains of the two grand potentials $\gpa$ and $\gpb$ 
coincide  by a scale transformation},   and {\em b)} {\em the state diagrams have a 
similar shape} in the sense that the 
state diagram  of system A converts into that of system B by the transformation 
(\ref{1.8})-(\ref{1.10}). [In fact, the state diagram of A is determined by the state 
variable points where the second  order partial derivatives of $\gpa$ are discontinuous. 
Equality (\ref{1.15}) and the assumed continuity of the partial second order derivatives of 
the state variable transformation imply a discontinuous behaviour 
of the relevant second order derivatives of $\gpb$.] Besides, the  TE  definition implies 
that: {\em c)} {\em if physical system A is \te\  to B and 
this is \te\ to system C, then systems A and C also are \te}.
 In fact, the TE  of\ B and C implies 
the existence of a reversible transformation: $V_C\to\Vb=V_{_{B,C}}$, 
$T_c\to\Tb=T_{_{B,C}}$ and $\mu_C\to\mub=\mu_{_{B,C}}$ [where $V_{_{B,C}}$, $T_{_{B,C}}$
and $\mu_{_{B,C}}$ denote functions that depend on $V_{C}$, $T_C$ and $\mu_C$]
such that 
\begeq\label{1.15b}
\gpb(V_{_{B,C}},T_{_{B,C}},\mu_{_{B,C}})=\Omega_{_{B,C}}(V_C,\,T_C,\,\mu_C)=
C_{_{BC}} \Omega_{{C}}(V_C,\,T_C,\,\mu_C).
\endeq 
On the other hand, the state variable transformation 
\begin{eqnarray}\label{1.15c}
 V_C\to\Va&=&V_{_{A,C}}\equiv V_{_{A,B}}(V_{_{B,C}},T_{_{B,C}},\mu_{_{B,C}}),\\ 
T_C\to\Ta&=&T_{_{A,C}}\equiv T_{_{A,B}}(V_{_{B,C}},T_{_{B,C}},\mu_{_{B,C}}),\label{1.15d}\\ 
\mu_C\to \mua&=&\mu_{_{A,C}}\equiv V_{_{A,B}}(V_{_{B,C}},\label{1.15e}
T_{_{B,C}},\mu_{_{B,C}})
\end{eqnarray}
 existes and is reversible. Applying this transformation to $\gpa$ and using equations 
(\ref{1.15}) and (\ref{1.15b}),  one gets 
\begin{eqnarray}\label{1.15f}
&&\gpa(V_{_{A,C}},T_{_{A,C}},\mu_{_{A,C}})=
\gpa(V_{_{A,B}},T_{_{A,B}},\mu_{_{A,B}})\Big|_{\Vb=V_{_{B,C}},\,\Tb=T_{_{B,C}},\,
\mub=\mu_{_{B,C}}}=\quad\quad\\
\ & &  C_{_{AB}}\gpb\bigl(\Vb,\,\Tb,\,\mub)\Big|_{\Vb=V_{_{B,C}},\Tb=T_{_{B,C}},
\mub=\mu_{_{B,C}}}=  C_{_{AB}} C_{_{BC}}  \Omega_{{C}}(V_C,\,T_C,\,\mu_C),
\nonumber
\end{eqnarray}
which proves the TE of A and C if A is \te\ to B and 
B to C.    This property implies that  the totality of the physical systems splits into 
different classes of  thermodynamically equivalent systems. \\ 
The TE  also implies that each thermodynamic quantity of system A is related to the 
corresponding quantity of system B in a known way.  Just to give an example, consider
 the entropies $\Sa$ and $\Sb$ of systems A and B. They are given by the relations 
$\Sa=-\partial \gpa(\Va,\,\Ta,\,\mua)/\partial \Ta$ and 
 $\Sb=-\partial \gpb(\Vb,\,\Tb,\,\mub)/\partial \Tb$. 
Taking the $\Tb$ derivative of equations (\ref{1.14a})-(\ref{1.15}) and recalling that  the 
pressure and  the system particle number are respectively given by
\begin{eqnarray}
p(V,\,T,\,\mu)&=&-\partial \Omega(V,\,T,\,\mu)/\partial V,\label{1.16}\\
 N(V,\,T,\,\mu)&=&-\partial \Omega(V,\,T,\,\mu)/\partial \mu,\label{1.17}
\end{eqnarray} 
 one finds 
\begin{eqnarray}
&&C_{_{AB}}\,S_B(\Vb,\, \Tb,\, \mub) =S_A (\Va,\, \Ta,\, \mua)
\frac{\partial {t_{a,b}}}{\partial\Tb}+\label{1.18}\\
\ &&\quad\quad \,p_A(\Va,\, \Ta,\, \mua)\Va\,\frac{\partial {v_{a,b}}}{\partial\Tb}+
\,N_A(\Va,\, \Ta,\, \mua)\frac{\partial {{\bar\mu}_{a,b}}}{\partial\Tb},\nonumber
\end{eqnarray}
so that the entropy of system B is a linear combination of the entropy, the pressure and 
the number particle of system A and the coefficients of the linear combination are known 
since they are appropriate  derivatives of the state variable transformation.  
The same property holds true for the pressure and  the system particle number. Similar 
inter-relations exist for other quantities as the specific heats at fixed particle number 
and at constant volume or at constant pressure, even though they become more involved 
since  these quantities are related  to higher order derivatives of the grand potentials.\\  
It is observed  that  the validity of equation (\ref{1.15})  is not sufficient to determine the 
coordinate transformation. In fact, if one tries to determine the transformation by a 
series expansion around a particular thermodynamic state, one realizes that the number of the 
involved unknown derivatives is larger  that the number of the equations. However if one 
identifies two of the state variables by setting, say,  $\Va=\Vb=V$ and $\Ta=\Tb=T$, { then 
equation (\ref{1.15}) might be sufficient to express $\mua$ in terms of $\mub$ and $T$ 
by a series expansion an, in this way, to determine the state variable transformation that 
converts $\gpa$ into $\gpb$.  It must be noted, however, that the resulting transformation 
as yet does not imply that two physical systems are thermodynamic equivalent. For this to happen,  one must check that the resulting transformation bi-injectively maps the state variable domain of A onto that of B.}\\ 
It is also noted that, in defining the TE, one has to choose a  thermodynamic potential.   One wonders if, assuming that 
two systems are TE with respect to, say, the grand potentials,  
they also are TE with respect to another thermodynamic potential. 
In section 3 it is reported a case where this happens, but the general answer is as yet 
unknown. \\ 
Finally,  the reported definition of TE could be made
less restrictive modifying condition (\ref{1.15}) as follows 
\begeq\label{1.20}\Omega_{_{A,B}}(\Vb,\,\Tb,\,\mub)=
\cC_{_{AB}}(\Tb,\mub)\gpb\bigl(\Vb,\,\Tb,\,\mub)+\cA_{_{AB}}(\Vb,\,\Tb,\,\mub),
\endeq
where $\cC_{_{AB}}$ and $\cA_{_{AB}}$ are  known functions of the reported state variables.  
{ This new definition, in contrast to that given by Eq. (\ref{1.16}), will be referred to as  
generalized thermodynamic equivalence (GTE). It obeys the transitive property  [\ie\ {\em c)}] but 
does not obey properties {\em a)} and {\em b)}. Unless one restricts the analytic form of 
$\cA_{AB}$, the GTE definition looks too much general to become nearly trivial. In fact,  
any invertible transformation of the state variables of systems A into those of B would yield the 
GTE of A and B by setting $\cA_{AB}=(\Omega_{AB}-\Omega_B)$ and $\cC_{AB}=1$.}  
For this reason the following analysis will  be mainly confined to definition (\ref{1.15}).  
\section{ The  hard rod gas in one dimension and the ideal classical gas in any dimension}
This section reports a first set of physical systems that are thermodynamically equivalent. 
They are the classical ideal gases in any space dimension $D$ and the hard rod gas in one 
dimension. Consider first  the pair formed by the one dimensional and the three dimensional 
classical ideal gases.  The grand potential  of the classical ideal gas, confined into  a box of 
length side $L$ and contained into
 a space of dimension $D(=1,2,3)$,  reads\refup{[1,2]}     
\begeq\label{3.1}
\gpigD(V_{_D},T,\muig)=-\kb T\, V_{_D}\,\frac { e^{\muig/\kb T}}{\lam^D},
\endeq 
with $V_{_D}=L^D$ and 
\begeq\label{3.4}
\lam\equiv h/\sqrt{2\pi m \kb T} 
\endeq 
denoting the de Broglie length,  $h$ and $\kb$  the Planck and the Boltzmann 
constants and $m$  the particle mass. 
For later reference it is convenient to report here also the expressions of the
 particle number density and the pressure, respectively given by 
\begeq\label {3.2}
\nig(T,\,\muig)= -\frac{1}{V_{_D}}\frac{\partial \gpigD(L,T,\muig)}{\partial \muig}=
\frac {e^{\muig/\kb T}}{\lam^D},
\endeq              
and  
\begeq\label {3.3}
\pig(T,\,\muig)= -\frac{\partial \gpigD(L,T,\muig)}{\partial V_{_D} }=
\kb\,T\,\nig(T,\,\muig).
\endeq          
To prove the TE of the considered ideal gases, one writes the 
right hand side of equation (\ref{3.1}) using definition (\ref{3.4}) as 
\begeq\label{3.4a}
-\frac{\kb (2\pi\,m)^{D/2}}{h^D} {T_{_D}}^{D/2+1}\, V_{_D}\,{ e^{\mu_{ig,\,D}/\kb T_{_D}}}.
\endeq
It is clear that the transformation
\begeq\label{3.4b}
T_1=T_3^{5/3},\quad  V_1=V_3/l_0^2,\quad{\rm and}\quad \mu_{ig,\,1}=\mu_{ig,\,3}T_3^{2/3},
\endeq
where $l_0$ denotes an arbitrarily chosen length, is invertible and continuous with 
all its partial derivatives. Once it is applied to $\gpigu(V_1,T_1,\mu_{ig,\,1})$ one 
finds that 
\begeq\label{3.4c}
\gpigu(V_1,T_1,\mu_{ig,\,1})=\frac{h^2}{2\pi\,m\,\kb\,l_0^2}\gpigt(V_3,T_3,\mu_{ig,3}),
\endeq 
which is equation (\ref{1.15}) with $C_{_{A,B}}=h^2/2\pi m\kb l_0^2$. Thus, the 
TE of the one-dimensional and the three-dimensional 
classical ideal gases is proved. By the same procedure one proves the  TE 
 of the one-dimensional and the two-dimensional 
ideal gases. In this way, by the transitiveness of the TE 
[see property {\em (c)} below equation (\ref{1.15})], 
one concludes that the $D$-dimensional ideal classical gases, whatever the 
positive integer $D$ value, form a class of  \te\  systems. \\
To prove that the one-dimensional ideal gas is \te\  to 
the one-dimensional hard rod gas of particles of mass $m$ and length $\sigma$, 
it is first recalled that the last system is characterized by the interaction potential 
$v_{_{hr}}(r)$ defined as   $v_{_{hr}}(r)=0$ if $r>\sigma$ and $v_{_{hr}}(r)=\infty$ 
if $0<r<\sigma$. 
Rayleigh\refup{[7]} and Tonk\refup{[8]} have since long determined the equation of  
state  of the one dimensional hard rod system. 
More recently Robledo and Rowlinson\refup{[9]} have determined the full 
thermodynamic behaviour of this system within the grand canonical framework  
paying great attention to the behaviour of the one particle distribution function 
close to the system walls.  The mentioned TE  between 
the ideal gas and the hard rod gas holds only true in the thermodynamic limit, 
\ie\ when the walls are infinitely far away. The grand partition function of the 
hard rod gas, confined to a box of size $L$, (after correcting a misprint 
present in the corresponding expression of reference {[9]} ) is 
\begeq\label{3.5}
\Xi_{hr}(L,T,\muhr)=\sum_{k=0}^{[L/\sigma]}\frac{{\zeta_{_{hr}}}^k}{k! }(L-k\,\sigma)^k.
\endeq 
with  
\begeq\label{3.6}
\zeta_{_{hr}} \equiv e^{\muhr/(\kb T)}{ \bigl/}\lam, 
\endeq 
and $[L/\sigma]$ equal to the integer part of $L/\sigma$.
The mean particle number of the particles $\left< N\right >_{_L}$ is 
\begeq\label{3.7}
\left< N\right >_{_L}=\zeta_{_{hr}}\frac{\Xi_{hr}(L-\sigma,T,\muhr)}{\Xi_{hr}(L,T,\muhr)}(L-
\sigma-\sigma\left< N\right >_{_{L-\sigma}}), 
\endeq
so that, in the limit  $L\to\infty$ with ${\left< N\right >_{_L}}/{L} = n_{hr}$ fixed [$n_{hr}$ 
denoting the bulk  particle number density of the hard rod fluid], one finds that 
\begeq\label{3.8}
\lim_{L\to\infty}\zeta_{_{hr}}\,\frac{\Xi_{hr}(L-\sigma,T,\muhr)}{\Xi_{hr}(L,T,\muhr)}= 
\frac{\nhr}{1-\sigma \nhr}.
\endeq
It also results that 
\begeq\label{3.9}
\frac{\partial \log{\Xi_{_{hr}}}}{\partial L} = 
\zeta_{_{hr}} \frac{\Xi_{hr}(L-\sigma,T,\muhr)}{\Xi_{hr}(L,T,\muhr)}.
\endeq
Using equation (\ref{3.8}) the above relation can be integrated to yield 
\begeq\label{3.10}
\log\bigl(\Xi_{_{hr}}(L,T,\muhr)\bigr)= \frac{L\,\nhr}{1-\sigma \nhr}+ const,
\endeq
which in turns implies that 
\begeq\nonumber
{\Xi_{hr}(L-\sigma,T,\muhr)}/{\Xi_{hr}(L,T,\muhr)}=
\exp\bigl(-\sigma\nhr/(1-\sigma\nhr)\bigr). 
\endeq
This result converts equation  (\ref{3.8})  into  
\begeq\label{3.11} 
\zeta_{hr} = \frac{\nhr}{1-\sigma \nhr}e^{\frac{\nhr\sigma}{1-\nhr\sigma}} 
\endeq 
that allows one to relate the chemical potential $\muhr$ to the particle number 
density $\nhr$. Putting $const=0$ in equation(\ref{3.10})  to ensure the 
extensiveness  of  the function, one finds that   the grand potential of the hard 
rod system is 
\begeq\label{3.12}
\gphr(L,T,\nhr)=-\kb T \log\bigl(\Xi_{hr}(L,T,\muhr)\bigr)= 
-\kb T \frac{L\,\nhr}{1-\sigma \nhr}.
\endeq 
This $\gphr$ expression depends on $\nhr$, while it is important to know $\gphr$ in 
terms of $\muhr$ in order to derive from it the other thermodynamic quantities. 
This can easily be realized using equation (\ref{3.11}) to express $\nhr$ in terms of 
$\muhr$.  To this aim one recalls that the solution, with respect to $x$, of the equation 
$y=xe^x$ is a transcendental function known as Lambert's function\refup{[10,11]} and  
denoted as $x=W(y)$.  This function also is the solution of the following differential 
equation
\begeq\label{3.13}
W'(y)=\frac{W(y)} {y(1+W(y))}=\frac{e^{-W(y)}}{1+W(y)}
\endeq
with the boundary condition $W(0)=0$.  Hence, the solution of  equation (\ref{3.11}) 
with respect to $\nhr$ is\refup{[11]}  
\begeq\label{3.14}
\nhr(T,\muhr)=\frac{W(\sigma e^{\muhr/\kb T}/\lam)}{\sigma\bigr(1+
W(\sigma e^{\muhr/\kb T}/\lam)\bigr)}.
\endeq 
Recalling that $W(y)\approx \log(y)$ as $y\to \infty$, this equation implies that 
$\nhr\to 1/\sigma$ as $\muhr\to\infty$ and $1/\sigma$ clearly represents the 
system density at the closest packing. 
The substitution of (\ref{3.14}) within equation (\ref{3.12}) yields the expression of 
the hard rod grand potential in terms of its natural variables, \ie
\begeq\label{3.15}
\gphr(L,T,\muhr)=-\kb T \frac{L}{\sigma}W(\sigma e^{\muhr/\kb T}/\lam).
\endeq
[The use of the same $\gphr$ symbol here and in equation (\ref{3.12}) is dictated by the sake 
of notational simplicity even  though it is not mathematically correct. This same convention 
will  be used  in a few of other points later.]  
The interaction of the hard rod fluid vanishes in the limit $\sigma\to 0$ so that the 
hard rod fluid  must behave as the ideal gas in the aforesaid limit. Since the Lambert 
function is such that $W(y)\approx y $ as $y\to 0$ one immediately realizes that 
\begeq\label{3.16}
\lim_{\sigma\to 0}\gphr(L,T,\muhr)=
-\kb T\, L\, e^{\muhr/\kb T}/\lam = \gpigu(L,T,\muig)
\endeq
once one sets $\muhr=\muig$, and the identical behaviour of the two systems in the 
limit $\sigma\to 0$ is proved. \\ 
The interesting point is that the hard rod  and the ideal gases are  \te\  
 if $\sigma >0$.  Consider, in fact, the following transformation 
of the state variables of the two systems 
\begin{eqnarray}\label{3.17}
\Tig=\Thr=T,&&\quad\quad \Lig=\Lhr=L\\
\muig=\muig(T,\muhr) &=& \kb T\Bigl(\log(\lam/\sigma)+
\log\bigl(W(\lam e^{\muhr/\kb T}/\sigma)\Bigr).\nonumber 
\end{eqnarray}               
It is a reversible transformation because the last equality can be inverted throughout 
the full range $(-\infty,\,\infty)$ of $\muhr$ to yield
\begeq\label{3.17a}
\muhr=\muhr(T,\muig)= \muig+\frac{\kb\, T}{\lam}e^{\frac{\muig}{\kb\,T}}.
\endeq
Once transformation (\ref{3.17}) is applied  to equation (\ref{3.1}) one finds that 
\begeq\label{3.18}
 \gpigu\bigl(L,T,\muig(T,\muhr)\bigr) =\gphr(L,T,\muhr),
\endeq 
that is equation (\ref{1.15}) with  $C_{_{AB}}=1$. In this way the TE  
 of the one dimensional hard rod and ideal gases is proved. Due 
to property {\em (c)}, one concludes that the D-dimensional ideal gases and the one 
dimensional hard rod gas form a class of \te\ systems. \\
The substitution of (\ref{3.17}) into equation (\ref{3.2}) and the use of equation 
(\ref{3.14}) imply  that the particle number densities of the two systems are related by   
\begeq\label{3.19}
\nig=\frac{\nhr}{1-\sigma \nhr}.
\endeq 
 This equation makes it evident that, as $\nhr$ ranges over the allowed physical 
range $[0,\,1/\sigma]$,  $\nig$ ranges over its physical range, namely the positive 
half-axis. The equation of state of the hard rod system is obtained by taking the 
derivative of $\gphr$ with respect to $(-L)$. It can directly be expressed in terms of 
variables $T$ and $\nhr$ derivating equation (\ref{3.12}). One finds
\begeq\label{3.20}
\phr=\kb T \frac{L\,\nhr}{1-\sigma \nhr}
\endeq
and, once one uses here relation (\ref{3.19}),  it coincides with that of the one-dimensional 
ideal gas $\pig=\kb T \nig$. \\ 
The entropy expressions are easily obtained from the grand potential expressions. 
For the hard rods fluid one finds 
\begin{eqnarray}\label{3.21}
S_{_{hr}}(L,T,\muhr)&=&\frac{L\kb}{\sigma} 
W(\sigma  e^{\muhr/(\kb T)}/\lam)  
\Bigl [ \frac{3}{2}+W(\sigma  e^{\muhr/(\kb T)} /\lam) -  \nonumber \\ 
\quad &&\quad\quad   \frac{\muhr}{\kb T}\,
\frac{1}{1+W(\sigma  e^{\muhr/(\kb T)} /\lam)}\Bigr],
  \end{eqnarray}
which, in terms of $\nhr$, reads
\begeq\label{3.22}
S_{_{hr}}(L,T,\nhr)= \kb\, L\, \nhr \Bigl(3/2  - 
    \log\bigl(\lam\,\nhr/(1 - \nhr \sigma)\bigr)\Bigr). 
\endeq
For the one-dimensional ideal gas one gets  
\begin{eqnarray}\label{3.23}
\Sig(L,T,\muig)&=& \kb\, L\, m\sqrt{\pi/2} e^{\muig/\kb T}\Bigl(3\kb T/2  -2\muig/(h 
\sqrt{\kb m T} )\Bigr)\quad  \nonumber  \\
\quad &=& \kb\, L\, \nig \Bigl(3/2 - \log(\lam\, \nig) \Bigr). 
\end{eqnarray}
The substitution of transformation (\ref{3.17}) into (\ref{3.23}) does not 
convert the ideal gas entropy in that of the hard rods. In fact, the  two entropies are related 
by the relation  [see equation (\ref{1.18})]
\begin{eqnarray}\label{3.24}
&&S_{_{hr}}(L,T,\muhr)=\Sig(L,T,\muig(T,\muhr))+N_{_{ig}}(T,\muig)
\frac{\partial\muig(T,\muhr)}{\partial T}, 
\end{eqnarray} 
because the $\muig$ transformation, given by (\ref{3.17}c),  depends also on T.\\
The expressions of the Helmotz free energies and internal energies 
of the two systems are now reported for completeness. The free energies are  
\begeq\label{3.27}
F_{_{hr}}(\Lhr, \Thr, \Nhr)= -\kb\, \Thr\,\Nhr\, \Bigl(1 - 
\log\bigl(\lam \nhr/(1 - \nhr \sigma)\bigr)\Bigr)
\endeq 
and 
\begeq\label{3.28}
F_{_{ig}}(\Lig, \Tig, \Nig)= -\kb\, \Tig\,\Nig \Bigl(1 -  \log\bigl(\lam \nig)\Bigr),
\endeq 
and the internal energies read 
\begeq\label{3.29}
U_{_{hr}}(\Lhr, \Shr, \Nhr)= \frac{h^2\,\Lhr}{4\pi\,m} \frac{ \nhr^3}{(1-\sigma\nhr)^2}\, 
e^{-3 + 2 S_{_{hr}}/(\kb \Nhr)}
\endeq
and 
\begeq\label{3.30} 
U_{_{ig}}(\Lig, \Sig, \Nig)= \frac{h^2\,\Lig}{4\pi\,m}\nig^3\, e^{-3 + 2 \Sig/(\kb \Nig)}.
\endeq  
One easily verifies that the  following  reversible transformation of the state variables
\begeq\label{3.31}
\Nig=\Nhr=N,\quad\Tig=\Thr=T,\quad \Lig=\Lig(\Lhr,N)=\Lhr -N\,\sigma
\endeq 
(with the bound $N<\Lhr/\sigma$)\ yields
\begeq\label{3.31}
F_{_{ig}}\bigl(\Lig(\Lhr,N), T, N\bigr)=F_{_{hr}}(\Lhr, T, N).
\endeq 
This relation proves the TE  of the two systems making 
use of  the Helmotz free energies. \\ 
If one considers, as thermodynamic potentials, the internal energies [see equations 
(\ref{3.29}) and (\ref{3.30})], it is straightforward to show that the following reversible 
state variable transformation 
\begeq\nonumber 
\Lig=\Lhr=L,\quad \nig=\frac{\nhr}{(1-\sigma\,\nhr)^{2/3}},\quad 
\Sig/\nig=\Shr/\nhr
\endeq 
yields 
\begeq\label{3.32}
U_{_{ig}}(\Lig, \Sig, \Nig)=U_{_{hr}}(\Lhr, \Shr, \Nhr),
\endeq
\ie\ the internal energies also are  \te.  
Equations (\ref{3.31}) and  (\ref{3.32}) suggest that if the grand potentials  are 
\te\ the same happens to the the other thermodynamic potentials,  
even though a general proof of this conjecture is still lacking. \\  
Before concluding the section, we report two properties of the hard rod fluid. The 
first concerns  the heat capacities  at fixed particle number and at fixed volume or 
at fixed pressure. The first  is easily derived from expression (\ref{3.22})  and the 
second from the same expression after expressing $\Lhr$ in terms of $\phr$  by  
equation (\ref{3.20}). In this way one respectively  finds 
\begeq\label{3.33}
C_{_{hr,V}}=\kb \Nhr/2\quad{\rm and}\quad C_{_{hr,p}}=3\kb \Nhr/2,
\endeq  
which  have the same analytic forms of the ideal gas heat capacities, \ie\ 
$C_{_{ig,V}}=\kb \Nig/2$ and $C_{_{ig,p}}=3\kb \Nig/2$. \\
The second property concerns the existence of the proportionality  relation 
between the grand potential and the internal energy, \ie 
\begeq\label{3.34}
\gphr(\Lhr ,\Shr,\Nhr)=-\frac{1}{1-\sigma\nhr}U_{_{hr}}(L,\Shr,\Nhr) 
\endeq
 This relation is obtained by substituting the temperature expression 
$T=-\partial U/\partial S$, resulting from equation (\ref{3.29}), into equation 
(\ref{3.12}) and then by comparing the result  with (\ref{3.29}).  
For the ideal gas the equivalent relation is\refup{[12]} is $\gpigu=
-\frac{1}{2}U_{_{ig}}$ that differs from (\ref{3.34}) because the proportionality 
coefficient is here strictly constant while it depends on the particle 
number density for the hard rod gas. 
\section{Systems having the same scaled interaction}
The systems considered in the previous section are characterized by interactions 
that coincide if the inter-particle distance exceeds $\sigma$.  In this section one considers 
the cases where the interactions have a similar behaviour throughout the full range of 
distances in so far they have  the form $gV(r/\sigma)$ with $V(r)$ fixed and $g$ and 
$\sigma$ variable. Hence, one assumes that system A is made up of  particles of mass 
$\ma$, diameter $\sigma_A$, 
and interaction   $g_A\,V(r/\sigma_A)$ and system B of particles of mass $\mb$, 
diameter $\sigma_B$ and  interaction  $g_B\,V(r/\sigma_B)$. The two systems are 
\te. In fact,  the classical grand canonical  partition 
function of system A, after integrating  over the particles' momentums, is
\begin{eqnarray}\label{2.1}
&&\Xi_A(\Va,\Ta,\mua)\equiv\sum_{k=0}^{\infty}\frac{\za^k \, {(2\pi \ma k_B\Ta)^k} }
{k! h^{3k}}\times \\
\ &&\quad\quad\quad \int_{\Va}\ldots \int_{\Va}
e^{-\frac{\ga}{k_B\Ta}\sum_{1\le i<j\le k}V(r_{i,j}/\siga)}d^{N} {\bf r}\nonumber
\end{eqnarray}
with 
\begeq\label{2.2}
\za\equiv e^{ \mua/k_B \Ta}.
\endeq
 The corresponding grand potential is 
\begeq\label{2.2a}
\gpa(\Va,\Ta,\mua)=-k_B\,\Ta\log\bigl(\Xi_{_{A}}(\Va,\Ta,\mua)\bigr).
\endeq
The grand partition function of system B is 
\begin{eqnarray}\label{2.3}
&&\Xi_B(\Vb,\Tb,\mub)\equiv\sum_{k=0}^{\infty}\frac{\zb^k \, {(2\pi \mb k_B\Tb)^k} }{k! h^{3k}}\times \\
\ &&\quad\quad\quad \int_{\Vb}\ldots \int_{\Vb}
e^{-\frac{\gb}{k_B\Tb}\sum_{1\le i<j\le k}V(r_{i,j}/\sigb)} d^{N} {\bf r}.
\end{eqnarray}
Putting $\sigab\equiv \siga/\sigb$ (and $\sigba\equiv \sigb/\siga$)  and 
performing the following change of the integration variables 
${\bf r}_i\to \sigab  {{\bf r}'}_j $,   the configuration integral present in the $k$th 
term of   series (\ref{2.1})
becomes                 
\begeq\label{2.4}
\sigab^{3k}\int_{\Va\sigba^3}\ldots \int_{\Va\sigba^3}
e^{-\frac{\ga}{k_B\Ta}\sum_{1\le i<j\le k}V({r'}_{i,j}/\sigb)} d^{N} {{\bf r}'}.
\endeq
At this point one easily verifies that the state variable transformation  
\begin{eqnarray}\label{2.5}
&&\Vb=\Vb(\Va)\equiv\Va{\sigba^3},\quad \Tb=\Tb(\Ta)\equiv g_B\,\Ta/g_A,\quad \\
&&\quad\quad {\rm and}\quad\zb=\za\frac{\ma^{3/2}\siga^3\gb^{3/2}}{\mb^{3/2}\sigb^3\ga^{3/2}},
\nonumber
\end{eqnarray}
the last being equivalent to  
\begeq\label{2.6}
\mub=\mub(\Ta,\mua)\equiv  \frac{\gb}{\ga}\Bigl[\mua + \frac{3}{2}k_B\,\Ta 
\log\Bigl(\frac{\ma\ga\siga^2}{\mb\,\gb\sigb^2}
\Bigr)\Bigr], 
\endeq
is reversible and yields 
\begeq\label{2.7}
\Xi_B\bigl(\Vb(\Va),\Tb(\Ta),\mub(\Ta,\mua)\bigr)\equiv \Xi_A(\Va,\Ta,\mua).
\endeq 
This, in turns, implies that 
\begeq\label{2.9}
\gpb\bigl(\Vb(\Va),\Tb(\Ta),\mub(\Ta,\mua)\bigr)= (\gb/\ga)\,\gpa(\Va,\Ta,\mua).
\endeq
In this way, according to  equation (\ref{1.15}) with $C_{_{BA}}\equiv \gb/\ga$, the TE 
 of systems A and B is proved. 
This result shows that a class of equivalent thermodynamic systems is formed by 
those systems such that  each of them is formed by particles, of a given mass and diameter, 
which interact with a potential having a particular coupling value and a fixed shape. 
Hence, the systems interacting with a Lennard-Jones potential form a class of 
\te\ systems. The same happens for the Coulombian 
systems.  It is noted that the systems, which are \te\ under 
the variable transformations (\ref{2.5}) and (\ref{2.6}), also obey the law of 
corresponding states\refup{[1]}. 
\section{The ideal quantum gases}
It is now shown that, within a space of dimension $D$,  the ideal classical gas is 
\te\ to the ideal Fermi one and is not \te\ to the 
ideal Bose one if one adopts equation (\ref{1.15}) as the TE definition.  
To this aim it is first recalled that Lee\refup{[3]} showed that  
the grand potential expressions  of the two ideal quantum gases can be 
expressed in terms of the same mathematical functions  as follows 
\begin{eqnarray}
\Omega_{_{B,D}}(\Vb,\Tb,\mu_{_{B}}) &=& -\Bigl[\frac{\gb\,\Vb\kb \Tb}{{\lam}^D}\Bigr]
{\rm Li}_{D/2+1}(\zb) , \label{4.1} \\
\Omega_{_{F,D}}(\Vf,\Tf,\mu_{_{F}})&=& \Bigl[\frac{\gf\,\Vf\,\kb \Tf}{{\lam}^D}\Bigr]
{\rm Li}_{D/2+1}(-\zf)\label{4.2}
\end{eqnarray} 
Here suffices $B$ and $F$ respectively refer to the Bose and the Fermi gas,  
 $\gb=2s_{_{B}}+1$ with $s_{_{B}}$ equal to the particle spin,  
$\lam=h/\sqrt{2\pi m \kb T}$, 
$\Vb=\Vf=L^D$ with $L$ equal to the length of the confining box edge and  
$D$ equal to the space dimensionality. Further,  $\zb(=e^{\mub/\kb T})$ and  
$\zf(=e^{\muf/\kb t})$ are the Bose and Fermi gas fugacities. Finally, 
${\Li}_{s}(z)$ is the so-called  polylogarithmic function\refup{[13]} of index $s$ that 
is simply related\refup{[14]} to $\Phi(z,s,a)$,  the Lerch function, by 
the  relation 
\begeq\label{4.3} 
 \Li_{s}(z)=z\,\Phi(z, s, 1),
\endeq 
so that the ideal Fermi and Bose grand potentials can also be expressed\refup{[15]} 
in terms of the Lerch function. This  is defined as\refup{[14]} 
\begeq\label{4.3b}
\Phi(z, s, a)\equiv \frac{1}{\Gamma(s)}\int_0^{\infty}
\frac{t^{s-1}\,e^{-a\,t}}{1-z\,e^{-t}}dt
\endeq
for ${\rm Re}(a)>0$, ${\rm Re}(s)>0$ and ${\rm Re}(z)<1$.  
It obeys the three properties
\begeq\label{4.4a}
\Phi(z, s, a)=\sum_{p=0}^{\infty}\frac{z^p}{(p+a)^s}\quad{\rm if} \quad |z|<1, 
\endeq
\begeq\label{4.4b}
\Phi(z, s-1, a)=\bigl(a+z\frac{\partial\ }{\partial z}\bigr)\Phi(z, s, a)
\endeq
and 
\begeq\label{4.5}
\Phi(z, s+1, a)=-\frac{1}{s}\frac{\partial\ }{\partial a}\Phi(z, s, a).
\endeq
The  integral expression (\ref{4.3b}) shows that the Lerch function is analytic in the complex $z$ plane 
cut along the real axis from one to $\infty$. Further, equation (\ref{4.3b}) shows that 
$\Phi(z, s, a)>0$ as $z$ ranges over $(-\infty,\,1)$. 
From this property and equation (\ref{4.3}) it follows that $\Li(z)>0$ if $0<\,z\,< 1$ and 
$\Li(z)<0$ if $z<0$.  Recalling that $\Li_{s}'(z)$, the derivative  of $\Li(z)$, obeys 
$\Li_{s}'(z)=\Li_{s-1}(z)/z$ [this relation immediately follows from equations (\ref{4.4b}) 
and (\ref{4.3})], it follows that  $\Li_{s}'(z)>0$ throughout $(-\infty,\,1)$. One concludes 
that $\Li_{s}(z)$ (with $s>1$)  monotonously increases as $z$ goes from $-\infty$ to 1.  
The behaviour of the $\Li_s(z)$s  for $s=\frac{3}{2},\, 2,\,\frac{5}{2}$, the  values associated 
to $D=1,\, 2$ and 3, are shown in Figure 1 for $-\infty<z<1$. Further, the leading behaviour   
of $\Li_s(z)$, as $z\to 1^-$,  is\refup{[13]} 
\begin{eqnarray}\label{4.6}
\Li_s(z)&\approx&\zeta(s) - \zeta(s-1) \delta + \frac{1}{2}
\bigl[\zeta(s-2) - \zeta(s-1)\bigr] \delta^2+ \\
\quad &&\delta^{s-1}\Gamma(1 - s) -  \frac{1}{2} (s-1)\delta^s\Gamma(1 - s)  ,\quad {\rm as}
\quad \delta\equiv(1-z)\to 0^+,\nonumber
\end{eqnarray}
where $\zeta(x)$ is the Riemann zeta function.  As $z\to \infty$,  $\Li_s(-z)$ behaves 
as\refup{[4,5]} 
\begin{eqnarray}
 \Li_s(-z)&\approx&-\frac{\log^s|z|}{\Gamma(1+s)} ,\quad {\rm as}\quad z\to  \infty.\label{4.7}
\end{eqnarray}
$\bigl[$This behaviour is obtained using  Joncqui\`ere's relation (see equation (1.11.1.16) of 
reference [14]) and the asymptotic expansion of the Hurwitz function $\zeta(a,b)$ as $b\to \infty$ 
in the complex plane (see equation (25.11.43) of  reference [10])$\bigr]$.\\ 
\begin{figure}[!h]
{\includegraphics[width=7.truecm]{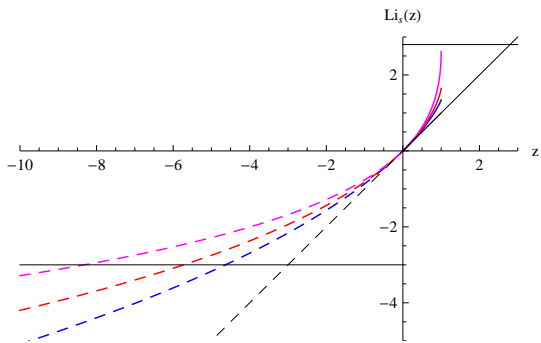}}
\caption{\label{Fig1} {The curves in colours are the plots of $\Li_s(z)$ for $s=5/2$ (blue), 
$s=2$ (red) and $s=3/2$ (magenta) within the range $-\infty<z<1$. The curves are broken 
within the negative $z$ half-axis and continuous within the range $0\le z<1$ to make
 it more evident that they respectively are the plots of $\Li_s(-\zf)$ with $\zf>0$ and 
of $\Li_s(\zb)$ 
with $0\le\zb<1$.  Note that the $\Li_s(z)$ limits are finite as $z\to 1$. The black broken 
and continuous straight lines are  the plots of $-\zig$ and $\zig$ as $\muig$ runs over 
$(-\infty,\, \infty)$.  The lower horizontal thin line shows that the equation 
$-\zig=\Li_s(-\zf)$ has a single root for each $\zig$ or $\zf$ value as well as for each 
of  the considered $s$ values. On the contrary, the upper horizontal thin line illustrates a 
case where the equation $\zig=\Li_s(\zb)$ has no roots.   
}}
\end{figure}
\noindent At this point, for each $D$ value,   the proof of the TE of the 
classical ideal gas and the ideal Fermi gas is straightforward. In fact, consider the state 
variable transformation 
\begeq\label{4.9}
\Tig=\Tf=T,\quad \Vig=\Vf=V\quad{\rm and}\quad  \zig=-\Li_{D/2+1}(-\zf),
\endeq 
the last relation being equivalent to 
\begeq\label{4.10}
\muig=\muig(\muf,T)\equiv \kb T\log\bigl(-\Li_{D/2+1}(-e^{\muf/\kb T})\bigr). 
\endeq
It is defined throughout the physical ranges of variables $\Tf$, $\Vf$ and $\zf$ and  
the  ranges of  the resulting variables $\Tig$, $\Vig$ and $\muig$ coincide with their  
relevant physical ranges. The transformations is also reversible. [This property, trivial for 
the first two variables, holds  true for variables $\zig$ and $\zf$ owing to the 
reported properties of the relevant $\Li_s(z)$ and it is also made evident by  
Figure 1.]  Then, the substitution of (\ref{4.9}) into (\ref{3.1})  and the comparison 
of the result with equation (\ref{4.2}) yield 
\begeq\label{4.11}
  \gpigD(V,T,\muig(\muf,T))=\frac{1}{\gf}\Omega_{_{F,D}}(V,T,\mu_{_{F}}),
\endeq 
that is condition (\ref{1.15}) with $C_{_{AB}}=1/\gf$. 
A corollary of this result is the property that, whatever the positive integer $D$, 
the ideal Fermi gases and the ideal classical gases as well as the one dimensional 
hard rod gas are all \te\ among themselves, \ie\ they 
form a class of \te\ systems. This property follows 
from the transitiveness property {\em (c)} and the TE of 
the D-dimensional ideal classical gases proved in section 3.\\ 
This result might look surprising because the classical ideal gas, in contrast 
with the Fermi one,  is physically inconsistent over a much wider range of 
the state variables as it happens, \eg, for the entropies and the specific 
heats. Nonetheless no contradiction is possible. In fact, 
it was already noted that the TE does not imply the equality of  corresponding 
thermodynamic quantities. Consider, for instance, the entropy case. The 
entropy of the Fermi gas is always positive\refup{[2]} while that of the ideal gas can be 
negative\refup{[16]}. Since they are related as in equation (\ref{1.18}), \ie 
\begeq\nonumber
S_{F}(V,T,\muf)=S_{ig}(T,V,\muig)+N_{ig}(T,\muig)\frac{\partial \muig(T,\muf)}
{\partial T},
\endeq
it follows that  the right hand side  turns out to be everywhere positive thanks to the 
contribution related to the partial $T$-derivative of  $\muig(T,\muf)$.\\ 
{ Finally, the ideal quantum Bose gases, compared to the ideal classical and quantum 
Fermi gases,  are now discussed from the point of view of the TE.  
Interestingly, the conclusion depends on the adopted 
definition of TE.  Consider first the restricted definition of TE given by Eq. (\ref{1.15}) 
and the ideal Bose gases with $D=1$ and $D=3$. Assume also that the temperatures  
and the volumes are related as in Eq. (\ref{3.4b}). Then  the expression  within the square 
brackets in Eq. (\ref{4.1}), in the case $D=1$,  becomes 
\begeq\label{4.11x}
\frac{\gb\,\Vb_1\kb \Tb_1}{{\lam_1}} \to \frac{\gb\,\Vb_3\kb \Tb_3}{{\lam_3}^3\,l_0^2},
\endeq 
that, aside for the factor $l_0^{-2}$, coincides with the expression relevant to the case 
$D=3$. The TE is ensured if one  proves the existence of an invertible transformation 
$\zb_1=\zb_1(\zb_3)$ with $0\le \zb_1 <1$ as $\zb_3$ ranges over $[0,\,1)$
such that 
\begeq\label{4.11y}
\Li_{3/2}\bigl(\zb_1(\zb_3)\bigr)   =\frac{\Li_{3/2}(1)}{\Li_{5/2}(1)}\Li_{5/2}\bigl(\zb_3\bigr). 
\endeq 
\begin{figure}[!h]
{\includegraphics[width=7.truecm]{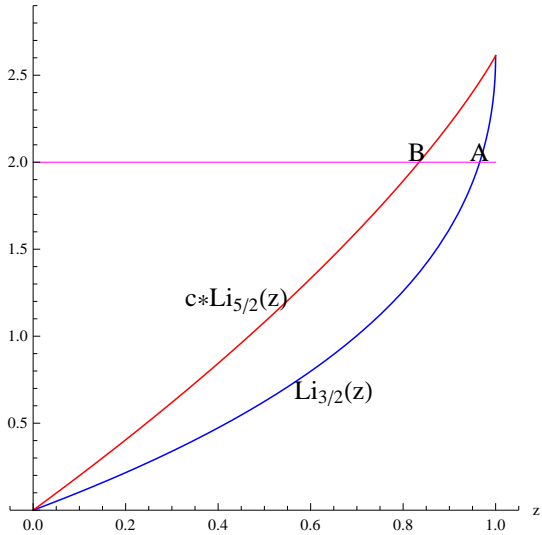}}
\caption{\label{Fig2} { The blue and the red curves respectively are the plots of $\Li_{3/2}(z)$ 
and $\Li_{3/2}(1)\Li_{5/2}(z)/\Li_{5/2}(1)$. Each horizontal line, of height 
smaller than $\Li_{3/2}(1)$  intersects the two curves at A and B. The abscissa of A 
is uniquely associated to that of B so as to define $\zb_1=\zb_1(\zb_3)$.}}
\end{figure}
Owing to the  reported properties of the polylogarithmic functions, the  above equation 
uniquely determines $\zb_1$ as function of $\zb_3$. Besides, the resulting function is 
continuous, endowed of continuos derivatives of any order, spans the interval $[0,\,1)$ 
and is  invertible within this interval.  Figure 2 makes these properties evident.  Substituting 
Eqs. (\ref{4.11x}) and (\ref{4.11y}) within (\ref{4.1}) one obtains 
\begeq\label{4.11z}
\Omega_{_{B,1}}(\Vb_1,T_1,\mu_{_{B,1}})=
 \frac{h^2\Li_{5/2}(1)}{l_0^2\,2\pi\,m\,\kb,\Li_{3/2}(1)}\Omega_{_{B,3}}(\Vb_3,T_3,\mu_{_{B,3}}),
\endeq    
where $\mu_{_{B,1}}=\kb {T_3}^{5/3}\log\Bigl(\bigl(\zb_1(e^{\mu_{_{B,3}}/\kb T_3}\bigr)\Bigr)$. 
Comparing (\ref{4.11z}) to (\ref{1.15}), one concludes that the ideal Bose gases in one 
and three dimensions are \te. The proof of the TE 
of the ideal Bose gases for the cases $D=2$ and $D=3$ proceeds quite similarly. 
Therefore, the ideal Bose gases, whatever the space dimension $D$,  form a class of 
thermodynamic equivalence. }\\ 
This class is different from that of the ideal Fermi gases and the 
classical ideal ones. To prove this statement it is sufficient to show that one physical 
system belonging to the first class is not TE to a system belonging to the second class.  
Consider then the ideal classical gas and the Bose quantum one in the case $D=3$ and  assume 
that the temperatures and the volumes of the two systems are equal. One  observes 
that, as $\zb$ ranges over the physical range $[0,1)$,  function $\Li_{5/2}(\zb)$ ranges 
over $[0, \zeta(5/2)\bigr)$ [see equation (\ref{4.6})]. Hence, at fixed $T$,  
$\Omega_{_{B,D}}(V,T,\mu_{_{B}})/\Bigl[\frac{\gb\,V\,\kb \,T}{{\lam}^D}\Bigr]$ 
ranges between $[0,\zeta(s)\bigr)$ as $\mub$ spans its physical 
range $(-\infty,\,0)$.  On the contrary, for the ideal gas, one finds that  
$\Omega_{_{ig}}(V,T,\muig)/\Bigl[\frac{V\,\kb T}{{\lam}^D}\Bigr]$ ranges 
over $(-\infty,\,0)$ as $\muig$ ranges over its physical range $(-\infty,\,\infty)$. 
The ranges of the two grand potentials are different and, therefore, the necessary 
condition   [see property {\em a)} of section 2] for  the two systems to be 
\te, in the restricted sense of Eq. (\ref{1.15}), is not 
obeyed. Thus, the proof of the non equivalence is achieved.  \\ 
{ This conclusion completely changes if one adopts the GTE definition  
given by Eq. (\ref{1.20}).  In fact, it will now be shown that, if one adopts the GTE definition,  
the aforesaid two distinct classes  of  restricted   \te\ systems 
combines into a single class of generalized \te\ systems. 
To this aim, it is first observed that the GTE implies the restricted TE and that transitive 
property {\em c)} of section 2 applies to both definitions. Then one recalls the 
result obtained by Lee\refup{[4]}, namely: the one-to-one fugacity transformation 
$\zf=\zf(\zb)\equiv\zb/(1-\zb)$,  in the case $D=2$,  relates 
the grand potentials and the one particle densities of the Bose and Fermi gases  
(outside the Bose condensation region so as to avoid Pathria's criticism\refup{[17]} and 
under the simplifying assumption that $\gb=\gf=1$)  as follows }
\begeq\label{4.12}
\Omega_{_{B,2}}(V,T,\mub)=\Omega_{_{F,2}}(V,T,\mu_{_{F}}(\mub)+
\frac{\kb\,T\,V}{2}\lam^2(T){\rho_{_B}}^2(\zb),
\endeq 
and 
\begeq\label{4.13}
\rho_{_B,2}(\zb)=\rho_{_F,2}(\zf(\zb))= \frac{1}{\lam^2}\Li_{1}(\zb). 
\endeq       
{The comparison of Eq. (\ref{1.20}) with (\ref{4.12}) and (\ref{4.13}) shows that the 
two-dimensional Fermi gas is GTE to the two-dimensional Bose gas with 
$\cA_{AB}(V,T,\mub)=\kb T V {\Li_1}^2(\zb)/\lambda^2$. In this way the GTE   
 of the classical ideal gases  as well as the  Fermi and the Bose quantum gases, 
whatever the space dimension $D$, is fully proved.  One should  also note that the added  
contribution $\kb T V {\Li_1}^2(\zb)/\lambda^2$ ranges ove $[0,\,\infty)$ as $\zb$ ranges over 
$[0,\, 1)$ so that the codomain of the right hand side becomes infinitely large as it happens 
to the function reported on the left hand side. }
\section{Conclusion} 
The consideration of the reversible transformations between the state variables 
of different physical systems naturally leads to the definition of thermodynamically 
equivalent systems. This paper mainly considered the most restrictive definition of the  
TE  which implies appropriate constraints on the codomains of the equivalent 
thermodynamic potentials as well as the similarity of 
the phase diagrams for the TE  to occur. The most interesting feature 
of the definition is the possibility of dividing all the physical systems into classes 
of TE. 
A first class of \te\ systems is formed by the ideal classical and 
quantum Fermi gases in any space dimensions and the one dimensional hard rod gas. 
A second class  is formed by the systems where the particle interaction can be brought 
to coincide by a scaling of the distance as well as of the coupling constant. Since these 
systems show phase transitions, they are therefore endowed of similar phase diagrams. 
{  A third class is formed by the ideal Bose gases in any spatial dimension and this 
class is different from the first one. But if one adopts the generalized definition of 
the TE the first and the third class become a single class of generalized TE.}
It is also stressed that   the TE of two different physical systems  does not ensure that  
all the thermodynamic quantities of one system are equal to the corresponding 
quantities   of the other system. It does only ensure that
 the quantities are linearly related among themselves with coefficients that depend 
on suitable (partial) derivatives of the state variable transformation. 
\section*{Acknowledgment}
I am grateful to Professor M. H. Lee for stimulating  correspondence and a critical reading of 
the manuscript.
\vfill\eject
\section*{References}
\begin{description}
\item[\refup{ 1}] A. M\"unster, {\em Statistical Thermodynamics}, Springer, Berlin (1969), Vol. I. 
\item[\refup{ 2}] L. D. Landau and E. Lifshitz,  {\em Physique Statistique}, Editions Mir, Moscou, (1967).
\item[\refup{ 3}] M. H. Lee, {\em J. Math. Phys.} {\bf  36}, 1217 􏰀(1995).
\item[\refup{ 4}] M. H. Lee, {\em Phys. Rev. E} {\bf 55}, 1518  􏰀(1997). 
\item[\refup{ 5}] M. H. Lee, {\em Acta Phys. Pol. B} {\bf 40}, 1279  􏰀(2009). 
\item[\refup{ 6}] D. V. Anghel, {\em J. Phys. A: Math. Gen.} {\bf 35}, 7255  􏰀(2002). 
\item[\refup{ 7}] L. Rayleigh, {\em Nature}, {\bf 45}, 80 (1891).
\item[\refup{ 8}] L. Tonks, {\em Phys. Rev.} {\bf 50}, 955 (1936).
\item[\refup{ 9}] A. Robledo and J.S. Rowlinson, {\em Molec. Phys.} {\bf 58}, 711  (1986).
\item[\refup{10}] {\em http://www.dlmf.nist.gov}
\item[\refup{11}] J.-M. Caillol, {\em J. Phys. A: Math. Gen.} {\bf 36},  10431,  (2003).
\item[\refup{12}] R. K. Pathria, {\em Statistical Mechanics}, Butterworth-Heinemann, Oxford (1996).
 \item[\refup{13}]  L. Lewin, {\em Polylogarithms and Associated Functions}, 􏰀North-Holland, 
New York, (1981)􏰁.
\item[\refup{14}] A. Erd\'elyi, W. Magnus, F. Oberhettinger and F. Tricomi, {\em Higher Transcendental Functions}, McGraw-Hill, New York, (1953), Vol. I .
\item[\refup{15}] S. Ciccariello, {\em J. Math. Phys.} {\bf  45}, 3353  (2004).
\item[\refup{16}] S. Ciccariello, {\em Eur. J. Phys.} {\bf  25}, 815  (2004).
\item[\refup{17}] R. K. Pathria, {\em Phys. Rev. E} {\bf 57}, 2697 􏰀(1998􏰁).

\end{description}
\end{document}